\documentclass[
 notitlepage,
 twocolumn,
superscriptaddress,
nofootinbib,
amsmath,amssymb,
aps,
pra,
longbibliography,
]{revtex4-2}
\pdfoutput=1
\usepackage[utf8]{inputenc}
\usepackage[english]{babel}
\usepackage[T1]{fontenc}
\usepackage{amsmath}

\usepackage{tikz}
\usepackage{soul}

\usepackage{amssymb}
\usepackage{graphicx}
\usepackage{physics}
\usepackage{pgf} 
\usepackage{color}
\usepackage{xcolor}

\usepackage[pdfstartview=XYZ,
bookmarks=true,
colorlinks=true,
linkcolor=blue,
urlcolor=blue,
citecolor=blue,
pdftex,
bookmarks=true,
linktocpage=true, 
hyperindex=true
]{hyperref}

\usepackage{orcidlink}

\newcommand{\ICFO}{ICFO - Institut de Ciencies Fotoniques, The Barcelona Institute of Science and Technology, 08860 Castelldefels, Barcelona, Spain}

\newcommand{\IOPPAS}{Institute of Physics PAS, Aleja Lotnikow 32/46, 02-668 Warszawa, Poland}

\newcommand{\UIBK}{Universität Innsbruck, Fakultät für Mathematik, Informatik und Physik, Institut für Experimentalphysik, 6020 Innsbruck, Austria}

\begin{document}

\title{
Exploring spin-squeezing in the Mott insulating regime: role of anisotropy, inhomogeneity and hole doping}

\author{Tanausú Hernández Yanes \,\orcidlink{0000-0001-8982-9277}}
\affiliation{\IOPPAS}
\email{hdez@ifpan.edu.pl}
\affiliation{\UIBK}
\author{Artur Niezgoda\,\orcidlink{0000-0002-2263-2725}}
\affiliation{\ICFO}
\author{Emilia Witkowska\,\orcidlink{0000-0003-4622-8513}}
\affiliation{\IOPPAS}

\begin{abstract}
Spin-squeezing in systems with single-particle control is a well-established resource of modern quantum technology. 
Applied in an optical lattice clock can reduce the statistical uncertainty of spectroscopic measurements.
Here, we consider dynamic generation of spin-squeezing with
ultra-cold bosonic atoms with two internal states loaded into an optical lattice in the strongly interacting regime as realized with state-of-the-art experiments using a quantum gas microscope.
We show that anisotropic interactions and inhomogeneous magnetic fields generate scalable spin-squeezing if their magnitudes are sufficiently small, but not negligible.
The effect of non-uniform filling caused by hole doping, non-zero temperature and external confinement 
is studied at a microscopic level demonstrating their limiting role in the dynamics and scaling of spin squeezing.
\end{abstract}

\maketitle

\section{Introduction}

Quantum technology is an emerging interdisciplinary field of study that combines the areas of physics, mathematics, and computer science. A prominent resource fueling emergent technologies like quantum simulators, computers and sensing is entanglement~\cite{PhysRevLett.96.010401, RevModPhys.89.035002, doi:10.1126/science.1097576,RevModPhys.90.035005}, a concept originating from the quantum mechanics formalism to explain correlations that cannot be reproduced classically. 
A plethora of useful entanglement-enhanced approaches are examined and spin squeezing is a well-established one~\cite{10.1063/5.0084096}. 

Spin-squeezing applies to a system composed of $N$ qubits, further described by the collective spin with the corresponding quantum number $S = N/2$. 
The uncertainty of spectroscopic measurements of unknown phase $\varphi$ for a given state is $\Delta \varphi= \xi/\sqrt{N}$, where
  \begin{equation}\label{eq:ssqparameter}
    \xi^2 = \frac{N\Delta^2\hat{S}_{\perp\min}}{\langle S\rangle^2},
\end{equation}
is the spin squeezing parameter while $\Delta^2\hat{S}_{\perp\min}$ is the minimal variance in the plane orthogonal to the direction of the mean collective spin $\langle S\rangle$ ~\cite{PhysRevA.46.R6797, PhysRevA.50.67}. 
If $\xi^2<1$, the corresponding state is spin-squeezed. However, a remarkable metrological gain is obtained with scalable spin squeezing when its level decreases significantly with the total number of spins. 

The archetypal model undergoing such desired scalability is the famous one-axis twisting (OAT) protocol (all-to-all interactions) where the best squeezing scales with the system size as $\xi^2_{\rm min}\propto N^{-2/3}$~\cite{PhysRevA.47.5138}.
It was simulated with pioneering experiments using bimodal Bose-Einstein~\cite{Treutlein2010,Oberthaler2010} and spinor~\cite{Chapman2012,PhysRevLett.125.033401,PhysRevLett.122.173601,Mao2023} condensates utilizing atom-atom collisions and atom-light interactions in cavity setups~\cite{PhysRevLett.104.073602,PhysRevLett.105.080403}. These platforms, however, weakly support a single-spin addressing and control required very often by quantum technology tasks.
There is an increasing interest in generation of spin-squeezed states using platforms where individual addressing of spins is possible~
\cite{Richerme2014, Kajtoch_2018, PhysRevA.102.013328, PhysRevResearch.1.033075,PhysRevLett.129.150503, PhysRevLett.129.113201,mamaev2023spin, PhysRevA.107.033318}. Recent experiments using an array of trapped ions~\cite{Franke2023} and Rydberg atoms~\cite{Bornet2023,Eckner2023} have demonstrated the generation of such scalable squeezing with tens of spins.  
Ultra-cold atoms in optical lattices offer yet another platform for scalable spin-squeezing generation in a system composed of tens of thousands of spins. 

In this paper, we study dynamical generation of scalable spin squeezing with ultra-cold bosonic atoms in two internal states loaded into a one-dimensional optical lattice.
We consider the strongly interacting regime with one atom per lattice site, where the system forms a ferromagnetic Heisenberg XXZ spin chain with nearest-neighbour interactions~\cite{Altman_2003, PhysRevLett.125.240504}. 
This is the Mott insulating regime.
The anisotropy of the corresponding XXZ spin model is set by intra- and inter-species interactions. 
When interaction strengths equal each other, the model reduces to the isotropic XXX Heisenberg spin chain.
We concentrate on the system and parameters as in the recent experiments with rubidium-87 atoms using the quantum gas microscope when the nearly single-atom control and resolution were obtained~ \cite{PhysRevX.9.041014}.
However, our analytical theory is general and can be applied to trapped ions and molecules when they simulate the same models~\cite{PhysRevA.108.052618,Micheli2006}.

Even for the simple system considered by us, experimental imperfections may arise such as slight anisotropy of the interactions, residual local magnetic fields, hole doping, external trapping effects or non-zero temperature.
They could negatively affect the dynamics of the system.
Throughout this work, we found that in most cases, not only these different effects can be accurately accounted for, but in most cases, they are beneficial for spin squeezing generation. 
  
We show analytically, and confirmed numerically, that a weak anisotropy of interactions allows generating scalable spin-squeezing from the initial spin coherent state, and the OAT model approximates well the dynamics. 
We evaluate analytically the time scale of the best squeezing showing its experimental feasibility. 
Adding a weak inhomogeneous magnetic field generates spin squeezing by itself, similarly. 
The coexistence of these two phenomena does not destroy squeezing generation but smoothly changes the time scale. 

We address the problem of hole doping on the generation of spin-squeezing.
We develop a microscopic theory to explain the change in the variation of the spin squeezing parameter in time due to hole doping. While the squeezing due to anisotropic interactions is weakly affected, the inhomogeneous field introduces sub-system rotations that modulate squeezing over time. 
This result is proven analytically for the case when holes are fixed in place.
We also explore the $t$--$J$ model, where tunnelling is allowed, and identify the upper and lower bounds for the generation of squeezing at a given filling factor.
We  find in the anisotropy case that squeezing immediately converges to the lower bound result if tunnelling is allowed. 
On the other hand, in the inhomogeneous magnetic field case the squeezing level hardly changes with the effective tunnelling. 
In both cases, the movement of holes facilitates the correlation between individual atoms initially belonging to different partial chains separated by these holes. 

The effect of harmonic trapping is also taken into account and even beneficial acceleration of dynamics is observed.
Lastly, we explore the effects of non-zero temperature on the squeezing generation of our model.
We illustrate our results for the parameters of experiments~\cite{PhysRevX.9.041014} demonstrating they can be realized with state-of-the-art techniques.

\section{Model}

We consider $N$ rubidium-87 atoms in two internal states 
$|a\rangle$ and $|b\rangle$ loaded in an optical lattice potential having $M$ lattice sites. For simplicity, we consider a one-dimensional lattice with open boundary conditions. The system is described by the two-component Bose-Hubbard model
\begin{equation}
\label{eq:BHM}
\begin{split}
\hat{\mathcal{H}}_{\rm BH} =& - J \sum\limits_{j, i=j\pm 1} \left(\hat{a}_{j}^{\dagger}\hat{a}_{i} + \hat{b}_{j}^{\dagger}\hat{b}_{i}\right) \\
&+ \frac{U_{aa}}{2}\sum\limits_{j} \hat{n}^a_j (\hat{n}^a_j -1)\\
& + \frac{U_{bb}}{2}\sum\limits_{j} \hat{n}^b_j (\hat{n}^b_j -1) 
+ U_{ab} \sum \limits_{j} \hat{n}^a_j\hat{n}^b_j ,
\end{split}
\end{equation}
in the lowest Bloch band and under the tight-binding approximation~\cite{RevModPhys.80.885}. $\hat{a}_j$ ($\hat{b}_j$) is the annihilation operator of an atom in internal state $a$ ($b$) in the $j$-th site of the lattice, and $\hat{n}^a_j=\hat{a}_{j}^{\dagger}\hat{a}_{j}$, $\hat{n}^b_j=\hat{b}_{j}^{\dagger}\hat{b}_{j}$ are the corresponding number operators. $J$ is the tunnelling rate, the same for bosons in the states $a$ and $b$. $U_{aa}$, $U_{bb}$ and $U_{ab}$ are specific intraspecies and interspecies interaction strengths. 
The model (\ref{eq:BHM}) can be realized using a quantum gas microscope~\cite{PhysRevX.9.041014, doi:10.1126/science.abk2397}. 
We assume interaction dominates over the tunnelling strength leaving the system in the Mott insulating regime. 
In the case of unit filling, one atom per lattice site, the effective Hamiltonian reduces to the Heisenberg XXZ model
\begin{equation}
\begin{split}
	\hat{H}_{\rm XXZ}=
    - J_\perp \sum_{j=1}^{M-1} 
    \bigg(& \hat{S}^x_{j} \hat{S}^x_{j+1} + \hat{S}^y_{j} \hat{S}^y_{j+1}  + \Delta \hat{S}^z_{j} \hat{S}^z_{j+1}  - \frac{1}{4} \bigg),
\end{split}
\label{eq:XXZ}
\end{equation}
where the couplings $J_\perp=4 J^2U^{-1}_{ab}$ and anisotropy parameter $\Delta=4 J^2 (U_a^{-1} + U_b^{-1} - U_{ab}^{-1})J_\perp ^{-1}$
are derived by second-order perturbation theory in the tunnelling~\cite{Altman_2003}. 
When $\Delta=1$ the Hamiltonian takes the form of isotropic Heisenberg XXX model.
Here, 
$\hat{S}^x_j = (\hat{S}^+_j+\hat{S}_j^-)/2$, 
$\hat{S}^y_j = (\hat{S}^+_j-\hat{S}_j^-)/(2i)$, 
$\hat{S}_j^z=(\hat{a}_j^\dagger\hat{a}_j - \hat{b}_j^\dagger\hat{b}_j)/2$ 
with $\hat{S}^+_j = \hat{a}_j^\dagger\hat{b}_j$, 
$\hat{S}^-_j=(\hat{S}^+_j)^\dagger$.
The collective spin operators are just a summation over the individual ones,
$\hat{S}_\sigma = \sum_{j=1}^M \hat{S}_j^\sigma$ for $\sigma = x,y,z, \pm$.

The generation of spin squeezing starts with the initial spin coherent state 
$|{\theta,\varphi}\rangle= e^{- i \varphi \hat{S}_z} e^{-i \theta\hat{S}_y }\bigotimes_{j=1}^M \ket{a}_j$ 
for $\varphi = 0$ and $\theta=\pi/2$ followed by unitary evolution with the Hamiltonian~(\ref{eq:XXZ}). Note, the state for $\varphi=\theta=0$ is the Dicke state $|S,m \rangle = \bigotimes_{j=1}^M \ket{a}_j$ for maximal spin quantum number $S=N/2$ and magnetization $m=N/2$. 
In our numerical simulations, we consider open boundary conditions~\cite{yanes2023spin} and use the parameters as in the recent experiment of A. Rubio-Abadal et al \cite{PhysRevX.9.041014} with $^{87}$Rb atoms, lattice spacing $d=532$ nm, tunneling amplitude $J = \hbar \times 2 \pi \times 24.8$Hz and almost equal inter- and intraspecies interactions $U_{aa} \sim U_{bb} \sim U_{ab}=U$~\cite{Fukuhara2013} with $U=24.4J$. For the sake of simplicity, time scales will be expressed in tunnelling units. The initial state is prepared as a coherent state along the $x$-direction in the Bloch sphere, namely $|{\theta=\pi/2,\varphi=0}\rangle$.
Finally, the spin squeezing parameter (\ref{eq:ssqparameter}) is evaluated for collective spin operators.

\section{Role of anisotropy}

\begin{figure}
    \centering
    \includegraphics[width=\columnwidth]{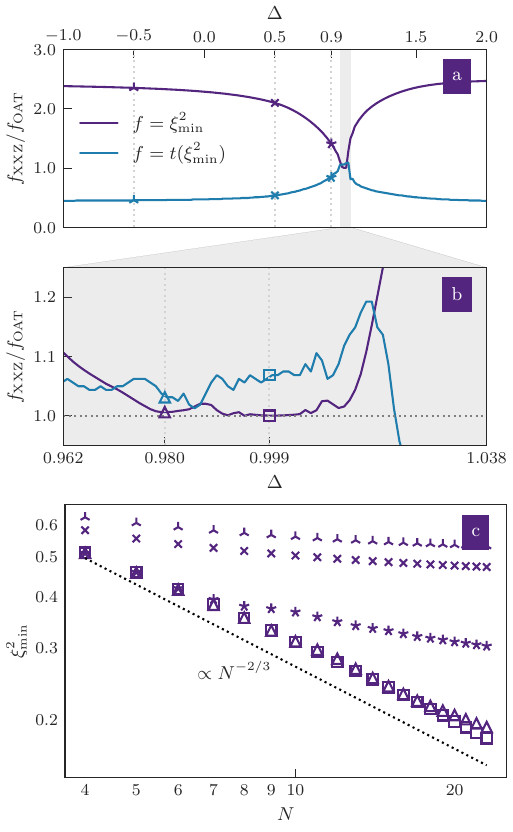}
    \caption{
    (a) The ratio $f$ between results of the XXZ model (\ref{eq:XXZ}) and the effective one (\ref{eq:eff0}) for 
    the best spin-squeezing $\xi^2_{\rm min}$ (first minima of $\xi^2$) and the best spin-squeezing time marked by purple and blue lines, respectively, when $M = N = 16$ and $\Delta \in (-1, 1)\cup (1, 2]$. 
    (b) The same as in (a) but in the perturbative regime enlarged for $\Delta \in [2\cos(\pi/N) - 1, 1) \cup (1, 3 - 2 \cos(\pi/N)]$.
    (c) The best spin-squeezing generated dynamically with the XXZ model (\ref{eq:XXZ}) for different values of $\Delta$ (see markers in upper and middle panels) versus $N$. A scalable level of squeezing is possible in the perturbative regime marked by triangles and squares.
    }
    \label{fig:fig1}
\end{figure}

The dynamical generation of spin squeezing is possible by anisotropic interactions, it is when $\Delta\ne 1$. We demonstrate this feature in Fig.~\ref{fig:fig1}(c) by plotting the best spin squeezing, see also Fig.~3 in \cite{mamaev2023spin}.
This numerical observation is confirmed by our perturbative analysis of the system Hamiltonian~(\ref{eq:XXZ}) when the term $\hat{H}_z = -J_\perp(\Delta-1) \sum_{j=1}^{M-1} \hat{S}_j^z\hat{S}_{j+1}^z$ is treated a perturbation to the isotropic XXX model.
This leads to the zero-order dominant term of the form
\begin{equation}
    \hat{H}^{(0)}_{\rm eff} = \chi_M^{(0)}\hat{S}_z^2, \,\,\, \,{\rm with}\,\,\,\,
    \chi_M^{(0)}=J_{\perp} \frac{1-\Delta}{M-1} ,
    \label{eq:eff0}
\end{equation}
where we omitted constant energy terms. Details of derivation are described in Appendix~\ref{app:effective-from-anisotropy}.
The resulting Hamiltonian (\ref{eq:eff0}) is the famous OAT model~\cite{PhysRevA.47.5138} which dynamics is solvable analytically for any $N$. The effective model approximates the dynamics of spin squeezing in the perturbative regime when $2\cos(\pi/M) - 1 \ll \Delta \ll 3 - 2 \cos(\pi/M)$ and $\Delta \ne 1$. In Appendix~\ref{app:OATSz} we collect specific analytical expressions for the first and second moments of spin components governed by OAT. 

The validity of our analytical finding is demonstrated in Fig.~\ref{fig:fig1}(a,b) by showing the relative level of best squeezing and best squeezing time obtained numerically from the full XXZ model (\ref{eq:XXZ}) and the effective one (\ref{eq:eff0}). For the considered set of parameters, the characteristic time scale for the best squeezing $\xi^2_{\rm min}$ is close to the one predicted by the OAT model, namely $J t/\hbar \simeq {3^{1/6}}(M-1)N^{-2/3} U_{ab}/(4 |1-\Delta| J)$. It is $J t/\hbar \simeq { 850}$ for  $M=N=16$,  $\Delta=0.98$ and $U_{ab} = 22.2 J (\Delta+1)$.
We found numerically for this set of parameters that the spin squeezing parameter for the OAT model reaches the minima at $Jt/\hbar \simeq 692$ while for the XXZ model at $Jt/\hbar \simeq 713$.

\section{Inhomogeneous magnetic field}

The addition of an external homogenous magnetic field $B \hat{S}_z$ to the XXZ model does not spoil spin-squeezing generation as long as $N$ is fixed. It contributes in the same form to the effective Hamiltonian~(\ref{eq:eff0}) leading to the model $\hat{H}_{\rm eff}= \chi_M^{(0)} \hat{S}_z^2 + B \hat{S}_z$ which dynamics is solvable analytically as shown in Appendix~\ref{app:OATSz}.

Similarly, even a weak inhomogeneous magnetic field 
\begin{equation}
\label{eq:HBinhomo}
\hat{H}_{\rm B} = \sum_{j=1}^M \beta_j \hat{S}_j^z .
\end{equation}
does not destroy spin-squeezing generation but changes the timescale of dynamics. To demonstrate this effect let us consider the isotropic case when $\Delta=1$ with an addition of a weak inhomogeneous magnetic field~(\ref{eq:HBinhomo}). 
The second-order correction obtained by using the Schrieffer–Wolff~(SW) transformation~\cite{Hern_ndez_Yanes_2022} takes the OAT form 
\begin{equation}\label{eq:Heff2}
    \hat{H}_{\rm eff}^{(2)} = - \chi^{(2)}_M \hat{S}_z^2 + v_M \hat{S}_z,
\end{equation}
when $\hat{H}_{\rm B}$ is treated as a perturbation to the XXX model and where
\begin{align}
    \chi^{(2)}_M &= \frac{1}{M-1} \sum_{q=1}^{M-1} \frac{|c_M^{(q)}|^2}{E^{(q)}_M},\\
    v_M &= \frac{1}{M}\sum_{j=1}^{M} \beta_j,
\end{align}
with 
\begin{align}
    c_M^{(q)}=&\sum_{j=1}^M p_{j}^{(q,M)} (\beta_{j} - v_M),\\
    p^{(q,M)}_j =& \sqrt{\frac{2}{M}} \cos\left[\frac{\pi}{M}\left( j -\frac{1}{2}\right)q \right],\\
    E_{M}^{(q)} =& J_{\perp} \left[1 - \cos (\frac{\pi}{M} q)\right],
\end{align}
for $q\in[1, M-1]$. 
The derivation is explained in Appendix \ref{app:effective-from-inhomo}. We omitted constant energy terms in (\ref{eq:Heff2}).

\begin{figure}
    \centering\includegraphics[width=\columnwidth]{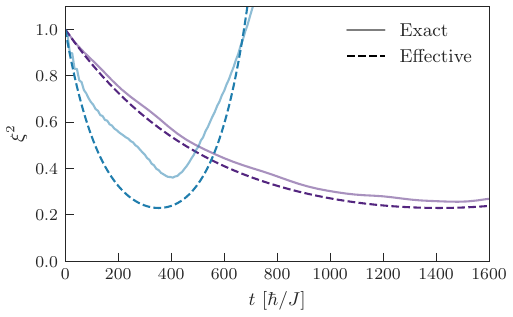}
    \caption{
    The spin squeezing parameter $\xi^2$ for isotropic XXX Heisenberg model with inhomogeneous field $\hat{H}_B$ and the effective one (\ref{eq:Heff2}) are marked by solid and dashed lines, respectively. 
    The amplitudes of magnetic field are $\beta_l = \sqrt{2/M} \cos{\left[\pi (l-1/2)(M-1) /M\right]}$.
    Here, $N = M = 16$, $J = 1, U_{aa} = U_{bb} = U_{ab} = U = 24.4 J, J_\perp = 4 \frac{J^2}{U}$, $\Omega = E_M^{(M-1)} / 10$ (purple lines) and $\Omega = E_M^{(M-1)} / 5$ (blue lines) with $ E_M^{(M-1)} = J_\perp (1 - \cos(\pi\frac{M-1}{M}))$. 
    }
    \label{fig:fig2}
\end{figure}
The validity of (\ref{eq:Heff2}) for the dynamical generation of spin squeezing via an inhomogeneous field (\ref{eq:HBinhomo}) when $\Delta=1$ is demonstrated in Fig.~\ref{fig:fig2}. This is an interesting example of when spin-squeezing, and therefore two-body correlations between elementary spins, are induced by inhomogeneity. In this case, the mechanism of spin-squeezing generation is caused by spin wave excitations which are extended over the entire system allowing individual spin to correlate~\cite{yanes2023spin}. It is the same mechanism as for the dynamical generation of spin squeezing via spin-orbit coupling~\cite{yanes2023spin, PhysRevResearch.1.033075}. 
This means any kind of inhomogeneous magnetic field (constant from shot to shot) will couple to the spin wave states, generating squeezing under the appropriate perturbation conditions.
We discuss this point in more detail in Appendix~\ref{app:effective-from-inhomo} where we also show other examples when the magnetic field takes a random value on each lattice site. This random potential also leads to the generation of two-body correlations and spin-squeezing in the perturbative regime.

\section{Doping of holes}

In this section, we consider an important effect coming from the non-occupied sites which we call holes.
In general, the dynamics is then captured by the $t$--$J$ model~\cite{RevModPhys.78.17,10.21468/SciPostPhys.5.6.057} which is the XXZ model with the additional tunnelling term. 
The tunnelling leads to the hole movement along the chain.
Here, we assume that positions of holes are fixed during unitary dynamics. The approximation is valid in the regime of parameters where the tunnelling is strongly suppressed as compared to $J_\perp$. 
The system dynamics can be then approximated by the XXZ model.
This allows an understanding of the role of holes at a microscopic level.

\begin{figure}
    \centering
    \includegraphics[width=\linewidth]{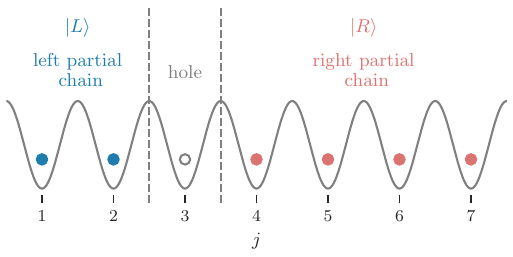}
    \caption{An example of a configuration with the position of a hole fixed. The hole located at the third lattice site $j_h=3$ separates the chain in two parts. The left(right) partial chain consists of two(four) spins. The two partial chains are independent Heisenberg spin chains with open boundary conditions. Their dynamics are independent of each other.}
    \label{fig:fig4}
\end{figure}

Let us first consider the simplest situation with one hole located somewhere in the middle of the chain (not at the borders) as illustrated in Fig.~\ref{fig:fig4}.
Since the hole is not moving, the configuration can be identified as two independent spin chains with open boundary conditions.

In this case, the collective Dick state for maximal magnetization containing the hole reads
\begin{equation}
|N/2, N/2 \rangle_{h} = | \uparrow \rangle_{j=1}\cdots 
| 0 \rangle_{j=j_h}\cdots | \uparrow \rangle_{j=M}.
\end{equation}
It can be represented as the product state of two partial Dicke states separated by the empty site,
\begin{equation}
\begin{split}
|N/2, N/2 \rangle_{h} =& | N_L/2, N_L/2\rangle \otimes | 0\rangle_{j_h} \\
&\otimes | N_R/2, N_R/2\rangle ,
\end{split}
\end{equation}
where $N_L$ is the number of spins on the left-hand side of the hole and $N_R$ is the number of spins on the right-hand side. The empty site does not contribute to the unitary dynamics driven by the XXZ Hamiltonian with nearest neighbours interactions. Therefore, we will omit the term $| 0\rangle_{j_h}$ when writing the states in the remaining part of the paper.

The initial spin coherent state for $\varphi = 0$ and $\theta=\pi/2$ reads
$|t=0\rangle_{h} = e^{-i \hat{S}_y \pi/2} |N/2, N/2 \rangle_{h}$.
It can be formulated in the following way:
\begin{equation}\label{eq:iniholes}
    |t=0\rangle_{h} = |L \rangle \otimes | R\rangle,
\end{equation}
where we have introduced
\begin{align}
    |L \rangle &= e^{-i \hat{S}_{y,L} \pi/2} |N_L/2, N_L/2 \rangle\\
    | R\rangle&= e^{-i \hat{S}_{y,R} \pi/2} | N_R/2, N_R/2\rangle ,
\end{align}
and used $\hat{S}_y = \hat{S}_{y,L} + \hat{S}_{y,R}$ with $L$($R$) summing up over the left(right)-hand part of the chain~\footnote{Here, 
$\hat{S}_{\sigma,L} =\sum_{j=1}^{N_L} \hat{S}_j^\sigma$ and 
$\hat{S}_{\sigma,R} =\sum_{j=1}^{N_R} \hat{S}_{N_L+1+j}^\sigma$.
}.

The dynamics of each partial chain (left and right) is independent of each other, and therefore the evolution of the initial state of the system can be considered as
\begin{equation}\label{eq:holeevol}
    |\psi(t)\rangle_h = \hat{U}_L |L \rangle \otimes \hat{U}_R | R\rangle,
\end{equation}
where the unitary operators are $\hat{U}_L=\hat{P}_L e^{i \hat{H}t/\hbar} \hat{P}_L$ and ($\hat{U}_R=\hat{P}_R e^{i \hat{H}t/\hbar} \hat{P}_R$) with $\hat{P}_L$ ($\hat{P}_R$) being the projector operator on the left (right) partial chain for a given Hamiltonian $\hat{H}$ containing nearest neighbours interactions only. 

The dynamics of partial chains is well approximated by effective OAT-like models for a weak anisotropy (\ref{eq:eff0}) and inhomogeneous magnetic field (\ref{eq:Heff2}) as we discussed in two previous sections. However, evolution operators acting on the left and right partial spin chains need to be constructed appropriately.

\subsection{Unitary evolution for partial chains for weak anisotropic interactions}
\label{subsec:LRani}

The effective model in the weak anisotropy limit when $2\cos(\pi/M) - 1 \ll \Delta \ll 3 - 2 \cos(\pi/M)$ and $\Delta \ne 1$ reads
\begin{equation}
    \hat{H}^{(0)}_{{\rm eff}, L} = -\chi_{L}^{(0)}  \hat{S}_{z,L}^2,
    \label{eq:eff0LR}
\end{equation}
for the left partial chain with $\chi_{L}^{(0)}=J_{\perp} (\Delta-1)/(N_L-1)$ and the same for the right partial chain when $L$ is replaced with $R$. 
Therefore, the unitary operator describing the dynamics with the hole for the left partial chains is
$\hat{U}_L= e^{- i \hat{H}^{(0)}_{{\rm eff, }L} t/\hbar}$ and similarly for the right partial chain when $L$ is replaced with $R$.

\subsection{Unitary evolution for partial chains for weak inhomogeneous magnetic fields}
\label{subsec:LRinho}

On the other hand, if spin squeezing is generated entirely by the inhomogeneous magnetic field (\ref{eq:HBinhomo}) for $\Delta=1$ the following effective model can well approximate the dynamics of the left partial chain
\begin{equation}\label{eq:HeffpartL}
	\hat{H}_{\rm eff, L} = \chi_L \hat{S}_{z,L}^2 + v_L \hat{S}_{z,L},
\end{equation}
with
\begin{align}
	\chi_L =& \frac{1}{N_L-1} \sum_{q=1}^{N_L-1} \frac{|c_{L}^{(q)}|^2}{E^{(q)}_{L}},\\
    v_L =& \frac{1}{N_L}\sum_{l=1}^{N_L} \beta_l,
\end{align}
where 
\begin{equation}
c_{L}^{(q)}= \frac{\sqrt{2}}{N_L}\sum_{l=1}^{N_L} p_{l}^{(q,L)} (\beta_{l}-v_L).
\end{equation}

The form of the effective model (\ref{eq:HeffpartL}) for the right partial chain is the same when one replaces $L$ with $R$ and where 
\begin{align}
v_R & = \frac{1}{N_R}\sum_{l=N_L+2}^{N_R} \beta_l\\
c_{R}^{(q)} &= \frac{\sqrt{2}}{N_R}\sum_{l=1}^{N_R} p_{l}^{(q,R)} (\beta_{N_L+1+l}-v_R).
\end{align}

The corresponding unitary operator describing the system dynamics of the initial state (\ref{eq:iniholes}) is 
$\hat{U}_{L(R)} = e^{- i \hat{H}_{\rm eff, L(R)} t/\hbar}$.

\subsection{Evaluation of spin squeezing parameter}

To calculate the evolution of the spin squeezing parameter (\ref{eq:ssqparameter}) one can use the approximated effective models as long as the system parameters are in the perturbative regime. This simplifies the calculations and enables the simulation of large systems unattainable by exact many-body numerical simulations.

To demonstrate the validity of our treatment of the system dynamics with hole doping, let us start with a general treatment of the first and second moments of spin operators that are necessary for calculations of $\xi^2$. 
The unitary evolution of first moments separates into two parts, e.g. if $\hat{X}=\hat{X}_L + \hat{X}_R$ we have
\begin{equation}
\label{eq:LRfirstmom}
    \langle  \hat{X} (t) \rangle_h = 
    \langle \hat{X}  \rangle_L + \langle \hat{X}  \rangle_R,
\end{equation}
where subscript $L$ ($R$) refers to the left (right) partial chain. $ \langle  \hat{X} (t) \rangle_h $ is a sum over the two partial chains, each evolved with the corresponding unitary operator. On the other hand, an expectation value of second moments is separated into four parts, e.g
$\hat{X} \hat{Y}= (\hat{X}_L + \hat{X}_R )(\hat{Y}_L + \hat{Y}_R)$, and reads
\begin{equation}
\label{eq:LRsecmom}
\begin{split}
    \langle \hat{X} \hat{Y} (t)\rangle_h =& \langle \hat{X}\hat{Y} \rangle_L + \langle \hat{X} \hat{Y} \rangle_R \\
    &+
    \langle \hat{X}\rangle_L \langle \hat{Y} \rangle_R
    +
    \langle \hat{X}\rangle_R \langle \hat{Y} \rangle_L .
\end{split}
\end{equation}
Each term in $\langle \hat{X} \hat{Y} (t)\rangle_h$ evolves with the unitary operator marked by the subscript $L$ or $R$.
While (\ref{eq:LRsecmom}) shows an apparent interconnection between partial chains, the covariance $\Delta(\hat{X}\hat{Y})_{h}^2 = \langle\hat{X}\hat{Y}\rangle_{h} - \langle\hat{X}\rangle_{h}\langle\hat{Y}\rangle_h$ turns out to be simply additive
\begin{equation}\label{eq:LRvariance}
    \Delta (\hat{X}\hat{Y})^2_{h} = \Delta (\hat{X}\hat{Y})^2_{L} + \Delta (\hat{X}\hat{Y})^2_{R}.
\end{equation}

The final ingredient to calculate $\xi^2$ is the mean collective spin $\ev{{S}}^2_h$. The breakage of the system into partial chains leads to the triangle inequality  $\langle {S}\rangle^2_h \le \langle{S}\rangle_L^2 + \langle{S}\rangle^2_R \le (N/2)^2$. This leads to 
\begin{equation}\label{eq:mean_spin_hole}
\ev{{S}}^2_h \le \left(\frac{N}{2}\right)^2,
\end{equation}
where equality can only happen when the hole is at one of the edges of the system. 

According to the definition (\ref{eq:ssqparameter}), this means that spin squeezing is immediately reduced when the system is broken into partial chains, however, the minimal variance can be optimal when the probability distributions of the partial chains add up appropriately.

\begin{figure}
    \centering
    \includegraphics[width=\linewidth]{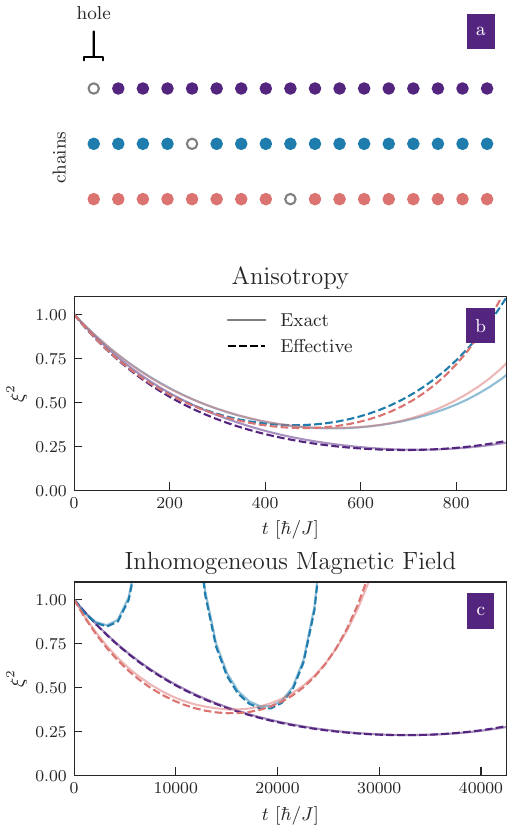}
    \caption{
    (a) Configuration of holes for results shown in bottom panels.
    The spin squeezing parameter induced by anisotropy (b) and inhomogeneous magnetic field (c) for the system with $N=16$ atoms and a single hole ($M=17$) at different sites as indicated by colour. 
    The results from the exact many-body simulation are depicted in solid lines while the ones given by the effective models are depicted in dashed lines, respectively.
    In the anisotropic case $U_{aa} = U_{bb} = 24.4J, U_{ab} = 0.99 U_a\ (\Delta = 0.98),\beta_j = 0; \forall j$.
    In the  inhomogeneous magnetic field case $U_a = U_b = U_{ab} = 24.4J, \beta_j = \Omega \cos(\frac{\pi}{M} (M-1)(j-1/2))$, where $\Omega = E_M^{(M-1)} / 50 = J_\perp (1-\cos\frac{\pi}{M}(M-1)) / 50$.
    }
    \label{fig:fig5}
\end{figure}

We illustrate this observation in Fig.~\ref{fig:fig5}. One can observe a good agreement between full many-body numerical calculations (solid lines) and approximated effective dynamics (dashed lines) as described in the subsections \ref{subsec:LRani} and \ref{subsec:LRinho}. 
In the anisotropy case, shown in Fig.~\ref{fig:fig5}(b), the partial chains obey the Hamiltonian (\ref{eq:eff0LR}) which maintains the mean spin direction across the short time dynamics. Thus, the main effect in the suppression of squeezing is due to the reduction of the collective mean spin. There is a secondary effect in the broadening of the minimal variance when the squeezing time scales of the partial chains differ (i.e. $\chi_{L}^{(0)} \neq \chi_R^{(0)}$).

\begin{figure}
    \centering
    \includegraphics[width=\linewidth]{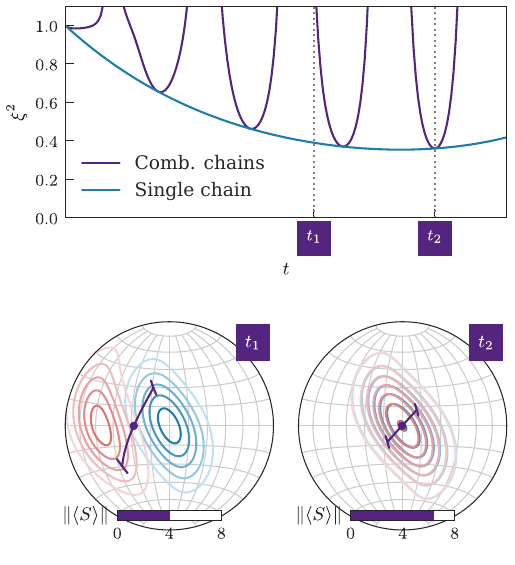}
    \caption{
    Modulation of spin squeezing dynamics for two partial chains, each with $N=8$, due to asynchronous rotation around the $\hat{S}_z$ axis. Both partial chains $L$ and $R$ obey the Hamiltonian (\ref{eq:HeffpartL}) where $\chi^{(0)}_L = \chi^{(0)}_R  =\chi$ but $v_R - v_L = 100\chi$. ($t_1$) The Husimi distribution with respect to coherent states of the partial chains ($Q(\theta,\phi) = |\braket{\psi_{L (R)}(t)}{\theta, \phi}|^2$) is drawn over the Bloch sphere in different colors at time $t_1$. The collective mean spin (lower bar) is reduced and the minimal variance (purple bracket) increases when the probability distributions separate.
    ($t_2$). At times $t \propto 2\pi/(v_R-v_L)$ the probability distributions overlap, maximizing the mean spin and reducing the minimal variance.
    }
    \label{fig:partial_chains_bloch}
\end{figure}

On the other hand, the results for spin squeezing in the presence of an inhomogeneous magnetic field (\ref{eq:HeffpartL}) includes a linear term that makes the probability distribution of each partial chain rotate around the $\hat{S}_z$ axis at different velocities. This creates oscillations in the squeezing parameter due to the misalignment of the partial mean spin directions, as illustrated in Fig.~\ref{fig:partial_chains_bloch}.
This feature is further discussed in Appendix~\ref{app:holes}.

\subsection{Generalization to any number of holes and configurations}\label{subsec:generalization}

The generalization of the above two-hole analysis to an arbitrary number of holes and their configurations is straightforward. 
In general, holes are located between occupied sites that constitute partial chains. 
All partial chains are independent as long as the positions of holes are fixed.

Let us start with the Dicke state for maximal magnetization written as the product state
\begin{equation}
\ket{N/2, N/2}_{\{h\}} = \bigotimes_{n} \ket{L_n/2, L_n/2} \bigotimes_{k\in \{h\}} \ket{0}_k,
\end{equation}
where $\{h\}$ describes the set of fixed locations of $N_h$ holes in the chain having $M$ sites, 
$n$ is the index numerating individual partial chains in the system and $L_n$ is the corresponding number of spins. The total number of spins in the whole chain is $N=M-N_h=\sum_n L_n$. We will omit the empty sites $ \ket{0}_k$ when describing the states in the further part of the text.

The initial spin coherent state is a product of coherent states of partial chains

\begin{equation}
\ket{t=0}_{\{h\}} = \bigotimes_{n} \ket{n},
\end{equation}
where
\begin{equation}
    \ket{n}=e^{-i \pi \hat{S}_{y,n}/2}\ket{L_n/2, L_n/2},
\end{equation}
for $\theta=\pi/2$ and $\phi=0$.

The further unitary dynamics is separable and each partial chain evolves independently 
\begin{equation}
\ket{\psi(t)}_{\{h\}}
= \bigotimes_{n} \hat{U}_n \ket{n},    
\end{equation}
meaning the state at any point in time can be described as a separable state of partial chains. 

In the case of spin squeezing generation by anisotropy, the Hamiltonian described in \ref{subsec:LRani} extends, and for each partial chain $n$ reads
\begin{equation}
    \hat{H}^{(0)}_{{\rm eff}, n} = -\chi_{n}^{(0)}  \hat{S}_{z,n}^2 ,
\end{equation}
with $\chi_{n}^{(0)}=J_{\perp} (\Delta-1)/(L_n-1)$.
When the spin squeezing is generated via the inhomogeneous field with $\Delta=1$, as discussed in subsection \ref{subsec:LRinho}, the effective OAT-like model for each partial chain is described effectively by the following Hamiltonian
\begin{equation}
	\hat{H}_{{\rm eff}, n} = \chi_{ n} 
  \hat{S}_{z,n}^2 + v_{n} \hat{S}_{z,n},
\end{equation}
where
\begin{align}
	\chi_{n}  &= \frac{1}{L_n-1} \sum_{q=1}^{L_n-1} \frac{|c^{(q)}_n|^2}{E^{(q)}_n},\\    v_n &= \frac{1}{L_n}\sum_{l=l_n}^{l_n+L_n-1} \beta_l,\\
    c^{(q)}_{n}&= \frac{\sqrt{2}}{L_n}\sum_{l=l_n}^{l_n+L_n-1} p_{l}^{(q,n)} (\beta_{l}-v_n),
\end{align}
where $l_n$ is the location of the first spin in the partial chain and $E^{(q)}_n = J_\perp (1-\cos(\pi q / L_n))$.

Finally, to calculate the spin squeezing parameter $\xi^2$ one needs to calculate the first and second moments of the the spin operators to obtain their covariances. 
Expectation values of an on-site linear operator can be described as a sum over all partial chains, $\ev*{\hat{X}}_{\{h\}} = \sum_n \ev*{\hat{X}}_{n}$ while for a product of two linear operators reads
$\ev*{\hat{X}\hat{Y}} _{\{h\}}=
\sum_n \ev*{\hat{X} \hat{Y}}_{n} + \sum_n \sum_{n'\ne n}\ev*{\hat{X}}_{n}\ev*{\hat{Y}}_{{n'}}$.
From these results one obtains $\Delta (\hat{X}\hat{Y})^2_{\{h\}} = \sum_n \Delta (\hat{X}\hat{Y})^2_{n}$.
The effective models for partial chains, as well as their separation, allow approximation of the dynamics of spin squeezing parameter by using analytical expressions shown in Appendix~\ref{app:OATSz} valid for any $N$ and $M$.

\subsection{Effective bounds when movement of holes is allowed}

In the previous subsections, we assumed the positions of particles and holes were fixed. However, a realistic scenario includes particle movement as stated in this section's beginning. The dynamics is then well captured by the $t$--$J$ model:
\begin{align}
\label{eq:t-J}
	&\hat{H}_{t-J}=- J \sum\limits_{i, j=i\pm 1} \hat{P}_0
 \left(\hat{a}_{i}^{\dagger}\hat{a}_{j} + \hat{b}_{i}^{\dagger}\hat{b}_{j}\right)\hat{P}_0
    + \hat{H}_{\rm XXZ} ,
\end{align}
where $\hat{P}_0$ is a projector operator over the manifold's ground states (i.e., single occupancy).
The evolution of a system is constrained to single occupied states but where particles can tunnel will be trivially bounded by two scenarios: no tunnelling and infinite tunnelling. 
The absence of tunnelling, $J=0$ in (\ref{eq:t-J}), is equivalent to a system where the holes are pinned down in fixed sites. 
Meanwhile, when tunnelling is effectively infinite, the system will behave as if fully occupied with a certain filling factor $f$ per site. 
Tunnelling is effectively infinite for a given time scale when the rest of the terms are energetically much smaller.
For instance, an increase in contact interactions in Eq. (\ref{eq:XXZ}) will in turn decrease $J_\perp$, increasing the time scale of the effective OAT model but making the tunnelling more prevalent. 

As a result, an increase in effective tunnelling will transform the evolution from the fixed holes scenario to the infinite tunnelling scenario.
We examined the scenario of fixed holes in previous sections.
The analytical result for the infinite tunnelling scenario is exactly the OAT model but expectation values are modulated by the filling factor. This is the lower bound for spin squeezing. In any intermediate cases with the holes tunnelling spin squeezing would be worse, depending on the energy scales ratio $J/J_\perp$.

To illustrate this process, we consider a simple statistical ensemble of $N_r$ realizations where the initial state has a fixed number of particles $N$ in a lattice of $M$ sites with $N < M$,
\begin{equation}\label{eq:rho_ensemble}
    \hat{\rho}(0)=
    \frac{1}{N_r} 
    \sum_{r =1}^{N_r} 
    \left[ \bigotimes_{j=1}^M 
    |\Psi_x\rangle \langle \Psi_x |_j
    \right],
\end{equation}
where 
over set of on-site random numbers $r\in \{x_1, x_2,  \cdots, x_{N_r}\}$.
Each state in (\ref{eq:rho_ensemble}) is 
\begin{equation}\label{eq:psi_delta}
    |\Psi_x\rangle_j = \bar{\theta}(f-x_j) |\uparrow \rangle_j 
    + \left( 1-\bar{\theta}(f-x_j) \right) |0\rangle_j,
\end{equation}
where $\bar{\theta}(x)$ is the Heaviside step function and $x_j\in (0,1]$ is an independent random number different for each lattice site. If $x_j\leq f$, then the site is occupied by an atom while if $x_j > f$ the site is empty resulting in a hole at that lattice site. In this way, we represented the presence of holes within the lattice by the filling factor $f$. 
Next, the state corresponding to each realization $r$ is rotated to form the spin coherent state for $\varphi = 0$ and $\theta=\pi/2$, $|t=0\rangle_{r} = e^{-i \hat{S}_y \pi/2} \left[ \bigotimes_{j=1}^M |\Psi_x\rangle \langle \Psi_x |_j \right]$, and unitary evolution is applied with the $t$--$J$ model.
To tackle the dynamics, we employ a semi-analytical approach. We analytically determine the dynamics of individual realisations using microscopic models developed in previous subsections but we treat the statistical ensemble numerically. This set the upper bound for spin squeezing at a given filling factor.

\begin{figure}
    \centering
    \includegraphics[width=\linewidth]{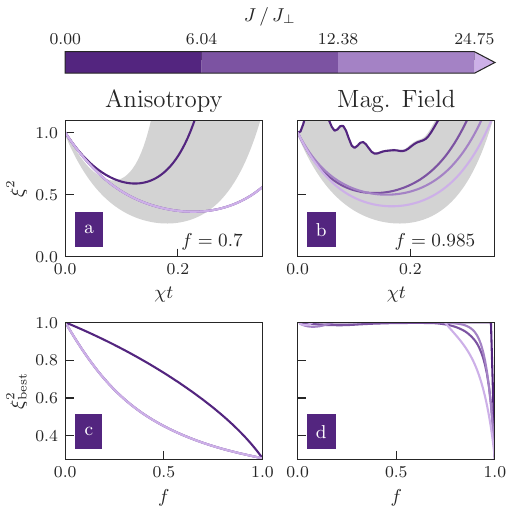}
    \caption{
    The evolution of spin squeezing parameter induced by anisotropy (a) and inhomogeneous magnetic field (b) for different values of the tunnelling $J$ and filling factors as indicated in the legend. 
    For each instance, $\chi$ is estimated as the corresponding parameter of effective models (\ref{eq:eff0}), (\ref{eq:Heff2}) when $N = M = 12$.
    Grey areas indicate the regions between the semi-analytical upper and lower bounds bounds as explained in the text. 
    The best squeezing $\xi^2_\mathrm{best}$ versus filling factor $f$ is shown in (c) and (d) when its generation is governed by anisotropy and inhomogeneous magnetic field, respectively.
    To tune the effective tunnelling, we fix $J=1$ but change $2 U_{ab} / (1+\Delta) \in \{24.4J, 50 J, 100 J\}$, with $J_\perp = J^2 / (4 U_{ab}), U_{aa} = U_{bb} = 2 U_{ab} / (1+\Delta)$. For the anisotropic case $\Delta = 0.98, \beta_j = 0$, while for the magnetic field case $\Delta = 1, \beta_j = E_M^{(M-1)} / 50 \cos{(\frac{\pi}{M}(M-1) (j-1/2))}$.
    }
    \label{fig:bounded_ensemble}
\end{figure}

In Fig. \ref{fig:bounded_ensemble} we compare the results for spin squeezing generation using anisotropy and inhomogeneous magnetic field when $M=12$. Since the effective model for the anisotropic case (\ref{eq:eff0}) lacks a linear term, the addition of multiple configurations of partial chains does not destroy squeezing and the inclusion of effective tunnelling immediately provides results close to the theoretical bound given by the OAT model. 
This contrasts with the inhomogeneous magnetic field case, where Eq. (\ref{eq:Heff2}). The presence of a linear term poses a challenge in achieving the infinite tunnelling limit. Since each configuration of holes returns a different velocity, we can picture an overlap between probability distributions as in Fig. \ref{fig:partial_chains_bloch}, but many of them move at different speeds.

\section{Effect of external confinement}

Up to now, we considered a homogeneous system with open boundary conditions. In this section, we show how the best squeezing time is tuned by an external harmonic potential without compromising the squeezing level, up to a certain threshold.
To be specific, we focus our attention on the simplest case with weak anisotropy, $\Delta\ne 1$, without a magnetic field or holes. 

\begin{figure}
    \centering
    \includegraphics[width=\linewidth]{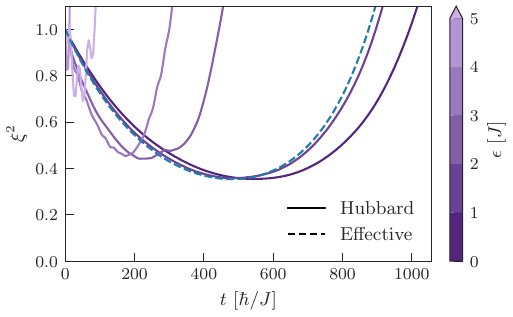}
    \caption{
    The effect of external harmonic trapping potential $\hat{V}_{\rm ext}$ on the spin squeezing dynamics obtained by numerical simulations from the two-component Bose-Hubbard model (\ref{eq:BHM}) in the Mott insulating phase compared to the effective model (\ref{eq:eff0}). $M = N = 8, J = 1, U_{aa} = U_{bb} = 24.4J, U_{ab} = 0.99 U_a \ (\Delta = 0.98)$. Solid line colours correspond to different values of $\epsilon$. The perturbation condition in this case is $\epsilon < 1.86J$.}
    \label{fig:fig3}
\end{figure}

The external trapping potential can be described in the second quantization form as
\begin{equation}
\hat{V}_\mathrm{ext} = \epsilon\sum_{j=1}^{M}\left(j-\frac{M+1}{2}\right)^2\left(\hat{n}_j^a + \hat{n}_j^b\right),    
\end{equation}
where $\epsilon = m \omega^2 / 2$ is the strength of the effective harmonic confinement with $m$ being the particle mass and $\omega$ the trapping frequency. Typically, the harmonic confinement is much smaller than the hopping rate, $\epsilon /J \lesssim 0.01$ \cite{doi:10.1126/science.1248402}.

In fact, as long as $J\ll U_{\sigma\sigma'}$, $\epsilon < \min\left(U_{\sigma\sigma}/2 - J, U_{ab} - J \right)/(M-2)$, double occupancy is unlikely and the effective OAT model~(\ref{eq:eff0}) well approximates the dynamics.  
In Fig.~\ref{fig:fig3} we illustrate the regimes above and bellow this threshold for a given set of parameters. For small values of $\epsilon$ the influence of the trapping potential is weak and only accelerates slightly the squeezing dynamics. On the other hand, a large $\epsilon$ means a promotion of double or larger occupied states, affecting the squeezing level as well.
In the limit case of trapping potential frequency of the order of individual lattice sites, the indistinguishably of individual spins is lost (all bosons located in a single lattice site), and our description is not valid. In this limit, the system is effectively bimodal. It is relevant to remark that at such a trapping strength, the system might actually also promote particles to higher bands outside the lattice \cite{PhysRevA.88.023620}.

\section{Non-zero temperature}

\begin{figure}
    \centering
    \includegraphics[width=\linewidth]{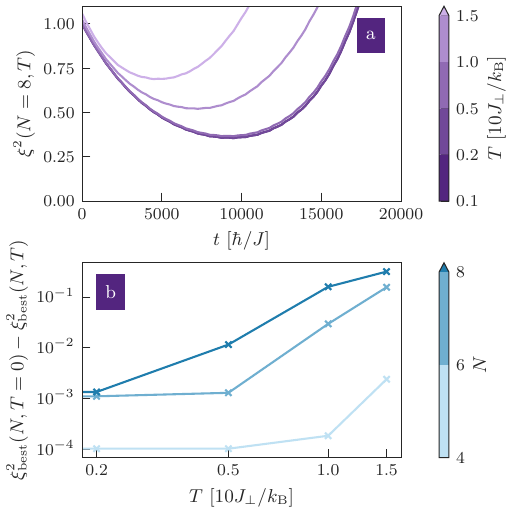}
    \caption{An illustration of the effect of non-zero temperature on the spin squeezing parameter. (a) The variation of the spin squeezing parameter in time for various temperatures is indicated in the legend. The smallest energy gap is 
    $J_\perp (1-\cos(2 \pi/N))\approx 0.29 J_\perp$ and $\Omega = 0.01 J_\perp$. (b) A difference between the best spin squeezing for zero and non-zero temperatures.
    }
    \label{fig:fig6}
\end{figure}

Thermal fluctuations limit the best squeezing achievable in two-component Bose-Einstein condensates~\cite{PhysRevLett.107.060404,Sinatra2012}. The same can be expected in the lattice system. To illustrate the effect we performed exact many-body numerical simulations.
We consider the isotropic Heisenberg XXX model, $\Delta=1$, exposed to the weak inhomogeneous field with
$\hat{H}_{\rm B} = \Omega/2\sum_{j=1}^M (e^{i \phi j} \hat{S}_j^+ + e^{-i \phi j} \hat{S}_j^-)$, $\phi=2 \pi/M$ and periodic boundary conditions as in~\cite{Hern_ndez_Yanes_2022}.
To observe spin squeezing, we choose the Gibbs state characterized by temperature $T$ as the initial state of the dynamics:
\begin{equation}
\label{eq:gibbs}
    \rho_T =\sum_{q=0}^{N-1} \frac{e^{-E_{q} /k_B T}}{Z} |q\rangle \langle q|
\end{equation}
next rotated with $\hat{R}=e^{-i \hat{S}_y \pi/2}$, namely,
\begin{equation}
    \hat{\rho}_R = \hat{R}^\dagger \hat{\rho}_T \hat{R} 
\end{equation}
to create the initial spin coherent state with $\varphi=0$ and $\theta=\pi/2$. In (\ref{eq:gibbs}) the ground state is the Dicke state $|q=0\rangle =  |N/2, N/2 \rangle$ with $E_0 = 0$, and higher energy states are spin wave states $\ket{q}$ given by
\begin{equation}
    |q\rangle = \frac{1}{\sqrt{N}}\sum_{l=1}^N e^{i q j 2 \pi/N} \hat{S}_l^- |N/2, N/2 \rangle,
\end{equation}
with periodic boundary conditions considered for this specific calculation, and where $q = 2\pi n/N$ and $n =
\pm 1, \pm 2,..., \pm(N/2 - 1), N/2$. The states $\ket{q}$ are eigenstates of
the total spin operator and its projection with the energy $E_q = J_{\perp} [1 - \cos ({2 \pi} q/N)]$. 
The thermally populated states are the lowest energy states of the $\hat{H}_{\rm XXX}$ Hamiltonian which are spin-wave states. The form (\ref{eq:gibbs}) is justified for $k_B T \ll  |E_{q=N/2}|$ when the temperature is much smaller than the largest energy gap when the occupations of the higher energy states are negligible. 

In Fig.~\ref{fig:fig6} we show numerical results for various temperatures. Admixture of higher energy states influences spin-squeezing dynamics and lowers the best squeezing generated in the system while the best squeezing time is shortened. However, as long as the temperature is much smaller than the smallest energy gap, $k_B T \ll E_{q=1}$ the effect is negligible as demonstrated in Fig.~\ref{fig:fig6}.
However, a detailed description of this effect goes beyond this work.

\section{Summary and conclusions}

We study the generation of scalable spin-squeezing with ultra-cold bosonic atoms in optical lattices in the Mott regime. 
This is possible through two main mechanisms related to imperfections in the system: anisotropy of contact interactions and inhomogeneous magnetic fields. 

We develop the microscopic theory to predict the dynamics of the spin squeezing parameter in the presence of hole doping in the simplified scenario when the positions of holes are fixed.
fIn the more general $t$--$J$ model, where a hole moves freely along the chain, the correlations in the system are bounded between the cases of zero and infinite effective tunnelling. The first case was considered by us in this paper at the microscopic level. In the second case, the movement of holes allows correlation of individual spins and, hence, the system behaves as fully occupied but the expectation values are modulated by the filling factor $f$. 
Additionally, we address numerically the question of the effect of external confinement and thermal fluctuations. While external trapping potential accelerates spin-squeezing dynamics, non-zero temperature diminishes the level of squeezing. However, in the latter case, the effect is negligible as long as the temperature value is much smaller than the smallest energy gap.

We believe our analysis sheds more light on the practical limitations of spin squeezing strategy for quantum technology tasks with ultra-cold atomic systems using a quantum gas microscope, or even trapped ions or molecules. However, we are aware that a transition from science to technology takes time and would happen when the quantum advantage outweighs the complexity of the experiments which are still under very extensive development. For example in the case of squeezed light, it took more than forty years for the successful application of entanglement-enhanced detection of gravitational waves~\cite{PhysRevLett.123.231107, PhysRevLett.123.231108}.

\section*{ACKNOWLEDGMENTS}
We gratefully acknowledge discussions with Bruno Laburthe-Tolra, Martin Robert-de-Saint-Vincent, Alice Sinatra, Youcef Bamaara and Manfred Mark. This work was supported by the Polish National Science Center DEC-2019/35/O/ST2/01873. 
T.H.Y acknowledges support from the Polish National Agency for Academic Exchange through the Foreign Doctoral Internship Grant NAWA Preludium BIS 1 No. PPN/STA/2021/1/00080. 
A part of the computations was carried out at the Centre of Informatics Tricity Academic Supercomputer \& Network.
A. N. was supported by Government of Spain (Severo Ochoa CEX2019-000910-S, TRANQI, European Union NextGenerationEU PRTR-C17.I1), European Union (PASQuanS2.1, 101113690) and the ERC AdG CERQUTE.

\begin{widetext}
\appendix

\section{Derivation of the XXZ model in the presence of anisotropy and inhomogeneous magnetic field}

We start our derivation from the two-component Bose-Hubbard Model for open boundary conditions with the addition of an inhomogeneous magnetic field

\begin{equation}
    \hat{H} = 
    \hat{\mathcal{H}}_{\rm BH} 
    + \hat{H}_{\rm B}, 
\end{equation}
where 
\begin{align}
\hat{\mathcal{H}}_{\rm BH} &= - J \sum\limits_{j, i=j\pm 1} \left(\hat{a}_{j}^{\dagger}\hat{a}_{i} + \hat{b}_{j}^{\dagger}\hat{b}_{i}\right) + \frac{U_{aa}}{2}\sum\limits_{j} \hat{n}^a_j (\hat{n}^a_j -1) 
+ \frac{U_{bb}}{2}\sum\limits_{j} \hat{n}^b_j (\hat{n}^b_j -1) 
+ U_{ab} \sum \limits_{j} \hat{n}^a_j\hat{n}^b_j ,\label{eq:BHM_app}\\
\hat{H}_{\rm B} &= \sum_{j=1}^M \beta_j \hat{S}_j^z. \label{eq:HBinhomo_app}
\end{align}

In fact, it does not have to be a magnetic field, it can be any other coupling that leads to the position-dependent external potential.
Notice $\hat{S}_j^z =  (\hat{n}_j^a - \hat{n}_j^b)/2$, so the local magnetic field is diagonal with respect to the Fock states.

The Bose-Hubbard Hamiltonian commutes with the total number of particles in each component
$\comm*{\hat{\mathcal{H}}_{\rm BH}}{\hat{N}_\sigma}=0$, where $\hat{N}_\sigma=\sum_j \hat{n}_{\sigma, j}$ with $\sigma = a,b$, but it does not commute with the occupation numbers $\hat{n}_{a, j},\, \hat{n}_{b, j}$ of the $j$-th site, due to the presence of the hoping terms. We address the case where the total filling is commensurate with the lattice.

We consider the system in the Mott phase when interaction dominates over the tunnelling strength. In the Mott regime, the system Hamiltonian is well described by the following model
\begin{align}
    \hat{H} = -\sum_{j=1}^{N-1} \left[
    J_{aa} \hat{n}_{j}^a  \hat{n}_{j+1}^{a} +
    J_{bb} \hat{n}_{j}^{b}  \hat{n}_{j+1}^{b}
    +J_{ab}^- \hat{n}_{j}^{a}  \hat{n}_{j+1}^{b}
    +J_{ab}^+ \hat{n}_{j}^{b}  \hat{n}_{j+1}^{a} 
    + J_\perp \frac{1}{2} \left( \hat{S}_j^+ \hat{S}_{j+1}^- + \hat{S}_j^- \hat{S}_{j+1}^+ \right)
     \right] + \hat{H}_{\rm B},
\end{align}
where
\begin{align}
    J_{aa}  & =\frac{4 J^2 U_{aa}}{U_{aa}^2 -  (\beta_j - \beta_{j+1})^2} \\
    J_{bb}  & =\frac{4 J^2 U_{bb}}{U_{bb}^2 -  (\beta_j - \beta_{j+1})^2} \\
    J_{ab}^- & =\frac{2 J^2 }{U_{ab} - (\beta_j - \beta_{j+1})} \\
    J_{ab}^+ & =\frac{2 J^2 }{U_{ab} + (\beta_j - \beta_{j+1})} \\
    J_\perp &= \frac{4 J^2 U_{ab}}{U_{ab}^2 - (\beta_j - \beta_{j+1})^2}
\end{align}
when taking into account the inhomogeneous field and after performing a SW transformation with the tunneling term as a perturbation.
The resulting system Hamiltonian can also be rephrased as
\begin{align}\label{eq:XXZ_full}
	\hat{H} =&- \sum_{j=1}^{N-1} 
	\left[ J_z \hat{S}^z_{j} \hat{S}^z_{j+1} + J_\perp \frac{1}{2} \left( \hat{S}_j^+ \hat{S}_{j+1}^- + \hat{S}_j^- \hat{S}_{j+1}^+ \right) - \frac{J_N}{4} \right] 
    + B \hat{S}_z + \sum_j h_j \hat{S}^z_j - \bar{h} \left(\hat{S}^z_1 - \hat{S}^z_N\right) + \hat{H}_{\rm B},
\end{align}
where
\begin{align}
    J_z & = J_{aa}+J_{bb}-J_{ab}^- - J_{ab}^+ , \\
    J_N &= J_{aa}+J_{bb}+J_{ab}^- + J_{ab}^+,\\
    B & = J_{bb} - J_{aa} ,\\
    h_j &= -\frac{J^2\left(\beta_j - \beta_{j+1}\right)}{U_{ab}^2-\frac{1}{4}\left(\beta_j - \beta_{j+1}\right)^2} +\frac{J^2\left(\beta_{j-1} - \beta_{j}\right)}{U_{ab}^2-\frac{1}{4}\left(\beta_{j-1} - \beta_{j}\right)^2},\\
    \bar{h} &= -\frac{J^2\left(\beta_N - \beta_1\right)}{U_{ab}^2-\frac{1}{4}\left(\beta_N - \beta_1\right)^2}.
\end{align}

We found numerically the influence of the $\bar{h}, h_j$ terms is negligible if the difference $\beta_j-\beta_{j+1} \ll U_{ab}$ and $\hat{H}_B$ dominates the perturbation of the XXZ model. In numerical calculations we will keep using Eq. (\ref{eq:XXZ_full}) but in further analysis we simplify the model in two main scenarios while also discarding these contributions and the homogeneous magnetic field $B \hat{S}_z$.

In the first case, we can take $\beta_j = 0;\ \forall j$, leading to the simple XXZ model in (\ref{eq:XXZ}). 

\begin{equation}
    \hat{H} = 
	\hat{H}_{\rm XXZ}=
    - J_\perp \sum_{j=1}^{M-1} 
    \left( \hat{S}^x_{j} \hat{S}^x_{j+1} + \hat{S}^y_{j} \hat{S}^y_{j+1}  + \Delta \hat{S}^z_{j} \hat{S}^z_{j+1}  - \frac{1}{4} \right),
\end{equation}
where  $J_\perp=4 J^2 / U_{ab}$ and the anisotropy parameter $\Delta=4 J^2 (U_{aa}^{-1} + U_{bb}^{-1} - U_{ab}^{-1}) / J_\perp$.
This can be further decomposed into an XXX model with perturbative term such that
\begin{equation}
    \hat{H} = \hat{H}_{\rm XXX} + \hat{H}_z,
\end{equation}
where
\begin{align}
	\hat{H}_{\rm XXX}= 
    &- J_\perp \sum_{j=1}^{M-1} 
    \left( \hat{S}^x_{j} \hat{S}^x_{j+1} + \hat{S}^y_{j} \hat{S}^y_{j+1}  + \hat{S}^z_{j} \hat{S}^z_{j+1}  - \frac{1}{4} \right),\label{eq:XXX_app}\\
    \hat{H}_z =& - J_\perp(\Delta-1) \sum_{j=1}^N \hat{S}_j^z\hat{S}_{j+1}^z.\label{eq:Hz_app}
\end{align}
The calculation of the resulting effective model is described in appendix \ref{app:effective-from-anisotropy}.

On the other hand, by choosing $U = U_{aa} = U_{bb} = U_{ab}$ and $U \gg (\beta_j - \beta_{j+1});\ \forall j$ one easily obtains
an XXX model with the inhomogeneous magnetic field.
\begin{align}
	\hat{H} = \hat{H}_{\rm XXX}  + \hat{H}_{\rm B}.
\end{align}
Excited states of the XXX model are given by the spin wave states \cite{yanes2023spin}, for which the $\hat{H}_{\rm B}$ term is a generator of. This leads to the effective model in (\ref{eq:Heff2}). See Appendix \ref{app:effective-from-inhomo} for more details.

\section{Derivation of the effective model from anisotropy}
\label{app:effective-from-anisotropy}

The initial state for unitary evolution is the phase state
\begin{equation}
    |{\theta,\varphi}\rangle= e^{- i \varphi \hat{S}_z} e^{-i \theta\hat{S}_y }\bigotimes_{j=1}^N \ket{a}_j
\end{equation}
which can be conveniently expressed in terms of the Dicke basis for maximal spin $S=N/2$, namely:
\begin{equation}
    |{\theta,\varphi}\rangle=\sum_{m=-N/2}^{N/2} \binom{N}{m+N/2}^{1/2}
    \left(  \cos{\theta/2}\right)^{N/2-m}  \left( e^{i \varphi} \sin{\theta/2} \right)^{N/2+m} \ket{m}.
\end{equation}
In the above representation $\ket{m}$ is the Dicke state as $\hat{S}^2 \ket{m} = S(S+1)\ket{m}$ and $\hat{S}_z \ket{m} = m\ket{m}$, and $\hat{S}^2$ and $\hat{S}_z$ are collective operators.

We consider the effective model describing the dynamics in the Dicke manifold where the initial state is localized. 
We derive the effective model in a perturbative way.
The unperturbed Hamiltonian is the XXX model (\ref{eq:XXX_app}) 
and it is weakly coupled to the anisotropy term $\hat{H}_z$ (\ref{eq:Hz_app}).

When the coupling is weak compared to the energy of the spin exchange $J_\perp$, the dynamics of the initial spin coherent state $|\theta, \varphi \rangle$ governed by the full Hamiltonian $\hat{H}=\hat{H}_{\rm XXX} + \hat{H}_z$ projected over the Dicke manifold can be well approximated using SW transformation~\cite{Hern_ndez_Yanes_2022, yanes2023spin} where the coupling $\hat{H}_z$ is treated as a perturbation. 

The dominant zero-order term $\hat{H}^{(0)}_{\rm eff}$ is determined by a projection of the coupling term over the Dick states and gives the following matrix representation:

\begin{equation}
    \langle m'|\hat{H}_z | m\rangle = -J_{\perp} (\Delta - 1) \left( - \frac{N}{4(N-1)} + \frac{m^2}{N-1}\right) \delta_{m', m}.
\end{equation}
Using the representation of the $\hat{S}_z$ operator we obtain
\begin{equation}
    \hat{H}^{(0)}_{\rm eff} = -J_{\perp} \frac{\Delta-1}{N-1} \hat{S}_z^2 + {\rm const}.
\end{equation}

\section{Derivation of the effective model from inhomogeneous magnetic field}
\label{app:effective-from-inhomo}

A weak inhomogeneous magnetic field
	$\hat{H}_{\rm B}$ (\ref{eq:HBinhomo_app}) 
can generate spin squeezing when added to the isotropic XXX Heisenberg model (\ref{eq:XXX_app}).
In fact it can be any other coupling which leads to the above form, also in different directions, e.g. $x$ or $y$.

To see this, one needs to calculate the second-order term $\hat{H}^{(2)}_{\rm eff}$ in perturbation, which matrix elements are defined as
\begin{equation}\label{eq:matrixheff}
\langle m'| \hat{H}^{(2)}_{\rm eff} |  m\rangle = 
- \sum_{m'',q} 
\frac{\langle { m'}|\hat{H}_{\rm B}|m'', q \rangle
	\langle m'', q |\hat{H}_{\rm B}| m\rangle}{E_{q}}.
\end{equation}
Details about the SW transformation and its application to the Heisenberg XXX model with the coupling can be found in~\cite{Hern_ndez_Yanes_2022}.
In the above equation states $\ket{m, q}$ are spin-wave states which are eigenstates of the isotropic Heisenberg model (\ref{eq:XXX_app}) for open boundary conditions~\cite{yanes2023spin}, namely
\begin{equation}
\label{eq:swsobc}
|m, {q} \rangle = \pm \sqrt{N} c_{N/2, \pm m} \sum_{j=1}^N p^{(q)}_j \hat{S}^{\pm}_j | m\mp 1\rangle,
\end{equation}
where
$c_{N/2, \pm m}=\sqrt{\frac{N-1}{(N/2\mp m)(N/2\mp m+1)}}$.
The sign $\pm$ in Eq.~\eqref{eq:swsobc} for $|m, {q} \rangle$ corresponds to two equivalent definitions of the spin waves in terms of the on-site spin raising and lowering operators $\hat{S}_j^\pm$ acting on the Dicke states $\ket{m}$. Furthermore, the coefficients featured in Eq.~\eqref{eq:swsobc} are
$p^{(q)}_j = \sqrt{\frac{2}{N}} \cos\left[\frac{\pi}{N}\left( j -\frac{1}{2}\right)q \right]$,
with $q=1,\cdots, N-1$. The corresponding eigenenergies $E_{q}$ of the isotropic model (\ref{eq:XXX_app}) read
$E_{q} = J_\perp \left[\cos (\frac{\pi}{N} q) - 1\right]$.

To calculate the form of the second-order term $\hat{H}^{(2)}_{\rm eff}$ it is useful to use the following commutation relations
$\left[ \hat{S}_{j}^{z}, \hat{S}_{-}^{n} \right]=-n \hat{S}_{j}^{-}\hat{S}_{-}^{n-1}$,
and
$\left[ \hat{S}_{j}^{z}\hat{S}_{j+1}^{z} , \hat{S}_{-}^{n} \right]=-n \left( \hat{S}_{j+1}^{-}\hat{S}_{-}^{n-1} \hat{S}_{j}^{z} +
\hat{S}_{j}^{-}\hat{S}_{-}^{n-1} \hat{S}_{j+1}^{z} \right) + n(n-1) \hat{S}_{j}^{-}\hat{S}_{j+1}^{-}\hat{S}_{-}^{n-2}$. 
They allow writing the action of $\hat{H}_{\rm B}$ on the Dick state $\ket{m}$ in the convenient form
\begin{equation}
\hat{H}_{\rm B}|m\rangle = \sum_{j}\beta_{j}\hat{S}_{j}^{z}|m\rangle=-\sqrt{\frac{S-m}{S+m+1}}\sum_{j}\beta_{j}\hat{S}_{j}^{-}|m+1\rangle ,
\end{equation}
to get the matrix elements in (\ref{eq:matrixheff}).

A term coupling directly the Dicke states  with the spin wave states of $q=0$ proportional to $\sum_j \beta_j$ will appear in $\bra{m'} \hat{H}_B \ket{m'',q}$ . However, $E_{q=0} = 0$, meaning we would have an infinite term.
To correct this, we simply have to make this sum zero by adding and subtracting a term to $\hat{H}_B$ so that

\begin{equation}
	\hat{H}_{\rm B}= \sum_{j=1}^{N} (\beta_j -v) \hat{S}^z_{j} + v \hat{S}_z = \hat{H}_{B'} + v \hat{S}_z,
\end{equation}
where $v = 1/N\sum_j \beta_j$.
This guarantees $\sum_j (\beta_j - v) = 0$, so if in the previous analysis we substitute $\hat{H}_B$ by $\hat{H}_{B'}$ we can correctly calculate (\ref{eq:matrixheff}).

The final expression for the effective Hamitonian $\hat{H}_{\rm eff}^{(2)}$ will then be
\begin{equation}
    \hat{H}_{\rm eff}^{(2)} = \chi \left( \hat{S}^2 - \hat{S}_z^2 \right) + v \hat{S}_z,\,\,\,\, {\rm where }\,\,\,\,
    \chi = \frac{1}{N-1} \sum_{q=1}^{N-1} \frac{|\sum_{l} p_{l}^{(q)} (\beta_{l}-v)|^2}{E_{q}}.
\end{equation}

\begin{figure}
    \centering
    \includegraphics[width=\columnwidth]{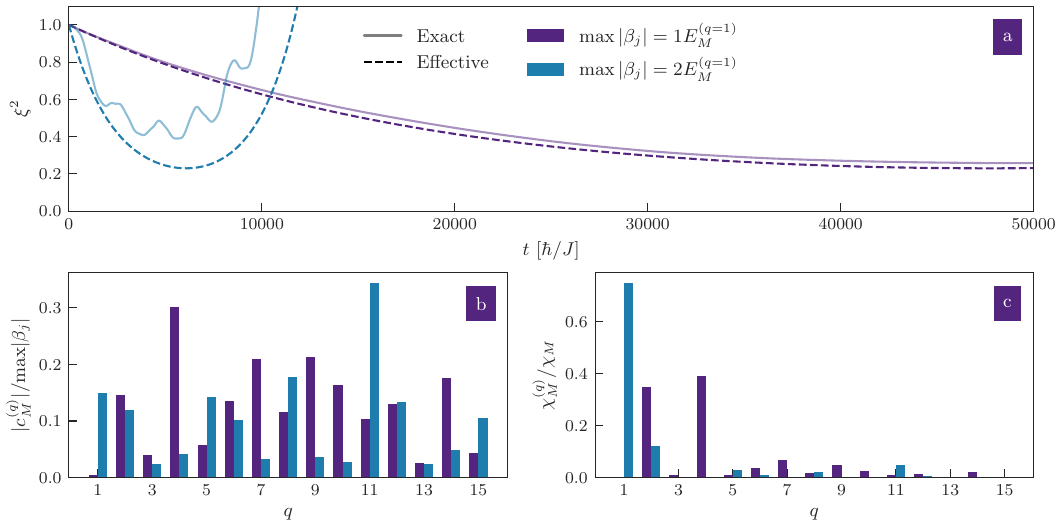}
    \caption{
    (a) Variation of spin squeezing parameter $\xi^2$ dynamically generated using the XXX model with a randomly generated inhomogeneous magnetic field (\ref{eq:XXX_app}) (solid lines) and the effective model (\ref{eq:Heff2}) (dashed lines). 
    Colors indicate the maximal magnitude of the magnetic field with respect to the smallest energy gap of the spin wave states. 
    Since $\chi_M \propto \max|\beta_j|^2$, the best squeezing time $t_\mathrm{best} \propto 1/|\chi_M|$ will be faster the larger this value is, in principle. 
    (b) The perturbation condition for each spin wave state is $\max|\beta_j| \ll E^{(q)}_M$. The effective model approximates the exact dynamics more accurately if the fidelity with spin wave states of quasi-momenta $q$ $\left(|c_M^{(q)}|/\max|\beta_j|\right)$ is negligible when the perturbation condition is not fulfilled (compare with (a)). 
    (c) Due to the functional form of the energy gap, the contributions of smaller $q$ in $\chi_M = \sum_q \chi^{(q)}_M$ tend to dominate the squeezing time scale.
    $N = M = 16$, $J = 1, U_{aa} = U_{bb} = U_{ab} = U = 24.4 J, J_\perp = 4 \frac{J^2}{U}, E^{(q)}_M =  J_\perp (1 - \cos(\pi q/M))$.
    }
    \label{fig:compare_sq_rand1}
\end{figure}

\section{Dynamics driven by $\chi S_z^2 + v S_z$}
\label{app:OATSz}
Consider the unitary evolution of the initial state
\begin{equation}
    | \Psi(t=0)\rangle = \sum_{m} c_m |S, m\rangle
\end{equation}
with the Hamiltonian
\begin{equation}
    H= \chi (\hat{S}^2 - \hat{S}_z^2) + v \hat{S}_z,
\end{equation}
namely:
\begin{equation}
    | \Psi(t)\rangle = \sum_{m} c_m e^{i \chi m^2 t - i v m t} |S, m\rangle
\end{equation}
where we omitted the constant phase factor.

One can express the evolution of spin operators in terms of evolution given by the pure OAT model when $v =0$.
Simple algebra shows that the first moments read
\begin{align}
    \langle \hat{S}_+ \rangle & = e^{i v t/\hbar} \langle \hat{S}_+ \rangle_{\rm OAT},\\
    \langle \hat{S}_- \rangle & = e^{-i v t/\hbar} \langle \hat{S}_+ \rangle_{\rm OAT},\\
    \langle \hat{S}_x \rangle & = \cos (v t/\hbar ) \langle \hat{S}_x \rangle_{\rm OAT} - \sin (v t/\hbar ) \langle \hat{S}_y \rangle_{\rm OAT},\\
    \langle \hat{S}_y \rangle & = \cos (v t/\hbar ) \langle \hat{S}_y \rangle_{\rm OAT} + \sin (v t/\hbar ) \langle \hat{S}_x \rangle_{\rm OAT},\\
    \langle \hat{S}_z \rangle & =  \langle \hat{S}_z \rangle_{\rm OAT} =0,
\end{align}
where for OAT we have
\begin{align}
    \langle \hat{S}_x \rangle_{\rm OAT}  &= S \cos ^{2 S -1} (\chi t/\hbar),\\
    \langle \hat{S}_y \rangle_{\rm OAT}  &=\langle \hat{S}_z \rangle_{\rm OAT}  =0.
\end{align}

On the other hand, the second moments are
\begin{align}
    \langle \hat{S}_+^2 \rangle & = e^{i 2 v t/\hbar} \langle \hat{S}_+ ^2 \rangle_{\rm OAT},\\
    \langle \hat{S}_-^2 \rangle & = e^{-i 2 v t/\hbar} \langle \hat{S}_-^2\rangle_{\rm OAT},\\
    \langle \hat{S}_+ \hat{S}_- \rangle & =
    \langle \hat{S}_+ \hat{S}_- \rangle_{\rm OAT},\\
    \langle \hat{S}_y \hat{S}_z \rangle & =
    \cos (v t/\hbar ) \langle \hat{S}_y \hat{S}_z \rangle_{\rm OAT} + \sin (v t/\hbar ) \langle \hat{S}_x \hat{S}_z \rangle_{\rm OAT}\nonumber\\
    &= \cos (v t/\hbar ) \langle \hat{S}_y \hat{S}_z \rangle_{\rm OAT},\\
    \langle \hat{S}_x \hat{S}_z \rangle & =
    \cos (v t/\hbar ) \langle \hat{S}_x \hat{S}_z \rangle_{\rm OAT} - \sin (v t/\hbar ) \langle \hat{S}_y \hat{S}_z \rangle_{\rm OAT} \nonumber\\
    &= - \sin (v t/\hbar ) \langle \hat{S}_y \hat{S}_z \rangle_{\rm OAT}, \\
    \langle \hat{S}_x^2 \rangle &=\frac{1}{2} \left(1+ \cos(2 v t) \right)\langle \hat{S}_x^2\rangle_{\rm OAT} 
    + \frac{1}{2} \left(1- \cos(2 v t) \right)\langle \hat{S}_y^2\rangle_{\rm OAT},\\
    \langle \hat{S}_y^2 \rangle &=\frac{1}{2} \left(1+ \cos(2 v t) \right)\langle \hat{S}_y^2\rangle_{\rm OAT} 
    + \frac{1}{2} \left(1- \cos(2 v t) \right)\langle \hat{S}_x^2\rangle_{\rm OAT} ,\\
    \langle \hat{S}_z^2 \rangle &=\langle \hat{S}_z^2 \rangle_{\rm OAT},\\
    \langle \hat{S}_x \hat{S}_y\rangle &= \cos 2 v t \langle \hat{S}_x \hat{S}_y\rangle_{\rm OAT } + \frac{1}{2}\sin 2 v t \left(\langle \hat{S}_x^2\rangle_{\rm OAT } - \langle \hat{S}_y^2\rangle_{\rm OAT }\right) \nonumber\\
    &= \frac{1}{2}\sin 2 v t  \left(\langle \hat{S}_x^2\rangle_{\rm OAT } - \langle \hat{S}_y^2\rangle_{\rm OAT }\right),
\end{align}
while the ones derived for the OAT model are
\begin{align}
    \langle S_x^2 \rangle_{\rm OAT} &= S/4 \left[ (2 S-1) \cos^{2 S -2} (2\chi t) +(2S+1) \right],\\
    \langle S_y^2 \rangle_{\rm OAT} &= - S/4 \left[ (2 S-1) \cos^{2 S -2} (2\chi t) - (2S+1) \right] ,\\
    \langle S_+S_- + S_-S_+ \rangle_{\rm OAT} &= 2 (\langle S_x^2 \rangle_{\rm OAT} + \langle S_y^2 \rangle_{\rm OAT}),\\
    \langle S_+^2 + S_-^2 \rangle_{\rm OAT} &= 2 (\langle S_x^2 \rangle_{\rm OAT} - \langle S_y^2 \rangle_{\rm OAT}),\\
    \langle S_+^2 - S_-^2 \rangle_{\rm OAT} &= 4 i \langle S_x S_y \rangle_{\rm OAT} = 0,\\
    \langle S_x S_z\rangle_{\rm OAT} &= 0 ,\\
    \langle S_z^2 \rangle_{\rm OAT} &= S/2,\\
    \langle S_y S_z\rangle_{\rm OAT} &= S (2 S -1)/2 \cos^{2S-2}(\chi t) \sin(\chi t).
\end{align}

\section{SWS with holes}
\label{app:holes}

One can construct the spin-wave states about these spin states separated by holes for maximal spin.
When considered in the Bethe basis:
\begin{equation}
	|l\rangle_h = \hat{S}_l^- |S=N/2, m= N/2 \rangle_{h}
\end{equation}
the spin wave states can be defined in the following way for the $N_h$ holes
\begin{equation}
	|n, q_n \rangle = 
	\sum_{l=1}^M p_l^{(q_n)}
  | l \rangle_{h},
\end{equation}
where $n$ numerates partial chains, and they are eigenstates of the $\hat{H}_{XXX}$ Hamiltonian when
\begin{align}
    p_l^{(q_n)} & = \sqrt{\frac{2}{L_n}}
    \cos \left[ \frac{\pi}{L_n} (l-(l_n-1/2)) q_n \right], \,\,\,\, l\in (l_n, l_n+L_n -1)\\
    p_l^{(q_n)} & = 0 \,\,\,\, \rm{otherwise}
\end{align}
with $l_n$ being the position of the first spin in the partial chain.

One can show that eigenenergies are:
\begin{equation}
    E_{q_n} = J_\perp \left[ 1 - \cos \left(\frac{\pi}{L_n} q_n\right)\right],
\end{equation}
where $L_n$ is the length of individual sub-chain (number of spins constituting the partial chains), while the corresponding quantum number of quasi-momentum $q_n \in [1, L_n -1]$.

Thus, $\hat{H}_{\rm XXX}+\hat{H}_\mathrm{B}$ leads through the second-order processes to the effective pure OAT model in each partial chain
\begin{equation}\label{eq:Heff}
	\hat{H}_{{\rm eff}, n}^{(2)} = - \chi_n \left( \hat{S}_n^2 - \hat{S}_{z,n}^2 \right) + v_n \hat{S}_{z,n},
\end{equation}
where
\begin{align*}
	\chi_n &= \frac{1}{L_n-1} \sum_{q_n=1}^{L_n-1} \frac{|c_{n}^{(q)}|^2}{E_{n}^{(q)}},\\
    c_{n}^{(q)}&= \frac{\sqrt{2}}{L_n}\sum_{l=l_n}^{l_n+L_n-1} p_{l}^{(q,n)} (\beta_{l} - v_n),\\
    v_n &= \frac{1}{L_n}\sum_{l=l_n}^{l_n+L_n-1} \beta_l,
\end{align*}
and one needs to set $l_n=1$ in $p_l^{(q,n)}$. Examples of various dynamics are presented in Fig.\ref{fig:comb_chains}.

\begin{figure*}
    \centering
    \includegraphics{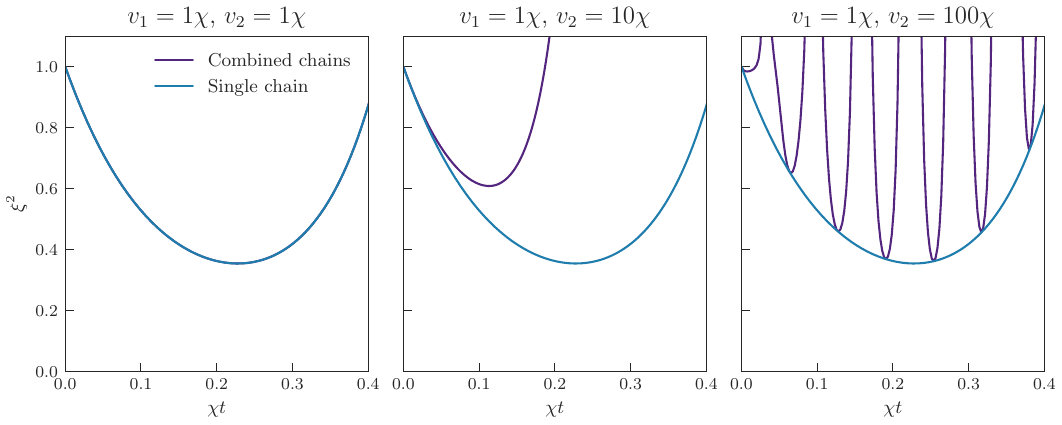}
    \caption{
    Comparison of spin squeezing parameter dynamics among combinations two of chains of $M=N=8$ particles with Hamiltonian (\ref{eq:Heff}) with same value of $\chi_n$ ($\chi_1 = \chi_2 = \chi$) but different values of $v_n$, indicated in the title of each panel.
    }
    \label{fig:comb_chains}
\end{figure*}

\end{widetext}

\bibliography{biblio}

\end{document}